\documentstyle[12pt]{article}
\def\x{\stackrel{\otimes}{,}}

\def\presentation{
\voffset -.50in
\hoffset -.19in
\oddsidemargin 0in \evensidemargin 0in
\marginparwidth .75in \marginparsep 7pt \topmargin 0in
\headheight 12pt \headsep .25in
\footheight 18pt \footskip .35in
\textheight 9.5in \textwidth 6.5in
\columnsep 10pt \columnseprule 0pt }

\presentation

\def\tilde{\widetilde}
\def\bar{\overline}
\def\hat{\widehat}
\def\*{\star}
\def\({\left(}          
\def\){\right)}         
\def\[{\left[}          \def\BBL{\Bigr[}
\def\]{\right]}         \def\BBR{\Bigr]}

\def\frac#1#2{{#1 \over #2}}            
\def\inv#1{{1 \over #1}}
\def\half{{1 \over 2}}
\def\d{\partial}

\def\ket#1{ | #1 \rangle}
\def\bra#1{ \langle #1 |}

\def\2pi{\hbox{$2\pi i$}}

\def\dsl{\raise.15ex\hbox{/}\kern-.57em\partial}
\def\Dsl{\,\raise.15ex\hbox{/}\mkern-.13.5mu D}
\def\ga{\gamma}

\def\ep{\epsilon}
        \def\La{\Lambda}
\def\de{\delta}         
\def\om{\omega}         \def\Om{\Omega}

       \def\CT{{\cal T}}

\def\debut{ \begin{eqnarray} }
\def\fin{ \end{eqnarray} }
\def\non{ \nonumber }

\begin{document}
\rightline{SPhT-93-072; LPTHE-93-40}
\vskip 1cm
\centerline{\LARGE The Sine-Gordon Solitons as a N-Body Problem.}
\vskip 1cm
\centerline{\large Olivier Babelon }
\centerline{Laboratoire de Physique Th\'eorique et Hautes
Energies \footnote[1]{\it Laboratoire associ\'e au CNRS.}}
\centerline{
 Universit\'e Pierre et Marie Curie, Tour 16 1$^{er}$
\'etage, 4
place Jussieu}
\centerline{75252 Paris cedex 05-France}
\vskip1cm
 \centerline{\large  Denis Bernard }
 \centerline{Service de Physique Th\'eorique de Saclay
\footnote[2]{\it Laboratoire de la Direction des Sciences de la
Mati\`ere du Commisariat \`a l'Energie Atomique.}}
\centerline{F-91191, Gif-sur-Yvette, France.}
 \vskip2cm
Abstract.\\
We consider the N-soliton solutions in the sine-Gordon model as a N-body
problem. This leads to a relativistic generalization of the Calogero model
first
introduced by Ruijsenaars. We show that the fundamental Poisson bracket of the
Lax matrix is quadratic, and the $r$-matrix is a dynamical one. This is in
contrast to the Calogero model where the fundamental Poisson bracket of the
Lax matrix is linear.
\newpage

\section{Introduction.}

We reconsider the relation between the sine-Gordon solitons and
the relativistically invariant N-body problem introduced by
Ruijsenaars. One of the motivations consists in extracting
coordinates suitable for quantization. Since in the quantum
theory, the asymptotic solitons are of primary importance
\cite{DAN76,ZaZa79},
soliton related coordinates seem appropriate. As illustrated by
eq.(\ref{EphQ}), the sine-Gordon field is expressed in a simple
way in terms of these coordinates. This kind
of formula could be useful for computing N-soliton contribution
to e.g. Casimir energy, form factors, etc...

The Ruijsenaars models \cite{Ruis} are relativistic generalizations of the
Calogero models \cite{Calo}. The later are integrable and even exactly solved:
the spectrum and more remarkably the wave functions are known.
Recently their rich algebraic structure began to appear \cite{Calsuite}.
It is tempting to believe that a similar structure will still exist
in these relativistic models.

The feature of the Calogero models that we want to extend
to the relativistic case is the existence of a $r$-matrix
\cite{AvTa93}. We find that the Poisson bracket
for the Lax matrix is quadratic, in constrast to the
non-relativistic case where it is linear.
The existence of  this quadratic Poisson bracket,
eq.(\ref{Ecata}), is probably a manifestation of the  quantum
affine symmetry of the sine-Gordon field theory \cite{Be90}.
Despite the fact that our motivation is the quantum theory,
in this Letter we restrict ourselves to the classical theory.
We illustrate the use of the $r$-matrix by providing simple
proofs for the commuting property of the Hamiltonians and
for the symplectic property for a remarkable coordinate
transformation.
It is interesting to notice that after quantization,
the Hamiltonians
eq.(\ref{EDa},\ref{EDb}) will lead to finite difference problems instead of
differential
equations.

We would like to point out that elliptic generalizations of these models exist.
It was
recently shown independently by Sklyanin, that a quadratic Poisson bracket also
exists
for their Lax matrix \cite{Skly}.

\section{Tau functions and solitons.}

We first introduce few notations for the sine-Gordon
equation and its solutions.
Let $z_\pm=x\pm t$ be the light cone coordinates and
$\d_\pm=\half(\d_x\pm\d_t)$. The sine-Gordon equation is:
\debut
\d_+\d_- \phi = 2 \sin(2\phi) \label{EAa}
\fin
It is convenient to introduce the two tau functions $\tau_\pm$,
which satisfy the Hirota equations:
\debut
\tau_\pm (\d_-\d_+ \tau_\pm )- (\d_+ \tau_\pm)( \d_-\tau_\pm)
= \tau_\pm^2 - \tau_\mp^2 \label{EAb}
\fin
The sine-Gordon field $\phi$ is related to the tau functions
by $\frac{\tau_+}{\tau_-} = \exp(-i\phi)$.

The tau functions of the $N$-soliton solutions of
the sine-Gordon equation are given by \cite{SCMc73,FaTa86,BaBe93}
\debut
\tau_\pm^{(N)}(z_+,z_-) = {\rm det}\(1\pm V\) \nonumber
\fin
with $V$ a $N\times N$ matrix with elements:
\debut
V_{ij}=2\frac{\sqrt{\mu_i\mu_j}}{\mu_i+\mu_j}\sqrt{X_iX_j}
\quad {\rm with}\quad   X_i=a_i\exp\({2\(\mu_iz_+ +\mu_i^{-1}z_-\)}\)
\label{EAc}
\fin
The quantities $a_i$ and $\mu_i$ are the parameters of the solitons:
$\mu_i$ are the rapidities and $a_i$ are related to the positions.
For the sine-Gordon equation, they satisfy specific reality conditions.
For solitons or antisolitons, the rapidity $\mu$ is real and $a$ is
purely imaginary, i.e. $a=i\ep e^\ga$  with $\ep=+1$ for a soliton
and $\ep=-1$ for an antisoliton. The ``breathers" correspond to
pairs of complex conjugated rapidities $(\mu,\bar \mu)$
and positions $(a,-\bar a)$. Notice that these conditions are
preserved by the dynamics.

The sine-Gordon equation is a Hamiltonian system. The symplectic form
is the canonical one:
\debut
\Om_{SG} = \int_{-\infty}^{+\infty} dx\ \de\pi(x) \wedge \de \phi(x) \nonumber
\fin
with $\pi(x)$ the momentum conjugated to the field $\phi(x)$.
Above, $\de$ denote the differential on the phase space.
One can take the restriction of this symplectic form to the $N$-soliton
subspace of the phase space. In the coordinates $a_i$ and $\mu_i$,
we find that this restriction is:
\debut
\om= \sum_{i=1}^N \frac{\de a_i}{a_i}\wedge \frac{\de \mu_i}{\mu_i}
+ \sum_{i<j} \({\frac{4\mu_i\mu_j}{\mu_i^2-\mu_j^2} }\)
\frac{\de \mu_i}{\mu_i}\wedge \frac{\de \mu_j}{\mu_j}
\label{EAd}
\fin
This two-form is non degenerate. It therefore defines
a symplectic structure on the restricted phase space.
The corresponding Poisson brackets are found by inverting the symplectic
form
(\ref{EAd}):
\debut
\{\mu_i,\mu_j\} &=& 0 \non \\
\{a_i,\mu_j\} &=& a_i \mu_j \de_{ij} \label{EAe}\\
\{a_i,a_j\} &=& - \({\frac{4\mu_i\mu_j}{\mu_i^2-\mu_j^2} }\)a_ia_j
\non \fin
How can Eq.(\ref{EAd}) be proved? First, we consider the one and two soliton
solutions. In these cases, the computation can be done directly
using the formula for the field $\phi(x)$ and the momentum
$\pi(x)=\d_t\phi(x)$. We find:
\debut
\Om_{SG}\Big\vert_{{\rm restricted}} = -2\int_{-\infty}^{+\infty} dx
\d_x\({\inv{\tau_+\tau_-}}\)\ \cdot\ \om \label{EAbr}
\fin
Thanks to the behaviour of the $\tau$-functions at infinity, the
prefactor is an irrelevant numerical constant.
Then, we consider the general case with an arbitrary number of
solitons with parameters $(\mu_i,a_i)$. Since
the dynamics is Hamiltonian, the symplectic form can be computed
at any time; in particular at $t\to \pm\infty$. In this in- or
out-limit, the sine-Gordon field becomes asymptotically
equal to the sum of the one-soliton  solutions
with parameters $(\mu_i^{\rm in} , a_i^{\rm in})$ and
$(\mu_i^{\rm out}, a_i^{\rm out})$ \cite{FaTa86}:
\debut
\mu_i^{{\rm in}} &=& \mu_i^{{\rm out}} = \mu_i \non\\
a_i^{{{\rm in }\atop {\rm out}}} &=& a_i \prod_{|\mu_j|{< \atop >} |\mu_i|}
\({\frac{\mu_i -\mu_j}{\mu_i+\mu_j} }\)^2 \nonumber
\fin
Since asymptotically the solitons decouple, in the symplectic
form the crossed terms vanish (the overlap integrals are zero).
Therefore the symplectic
form  reduces to the sum of the  one-soliton
expressions, but with the shifted in and out parameters:
\debut
\om=\sum_{i=1}^N \frac{\de a_i^{{\rm in }}}{a_i^{{\rm in }}}
\wedge \frac{\de \mu_i}{\mu_i}
=\sum_{i=1}^N \frac{\de a_i^{{\rm out}}}{a_i^{{ \rm out}}}
\wedge \frac{\de \mu_i}{\mu_i} \nonumber
\fin
This is equal to eq.(\ref{EAd}). If breathers are present the
proof is identical since we checked eq.(\ref{EAbr}) up to two
solitons.

As a by product, we see that the
transformation from the in-variables $(\mu_i^{{\rm in}},a_i^{{\rm in}})$
to the out-variables $(\mu_i^{{\rm out}},a_i^{{\rm out}})$
is symplectic. The classical $S$-matrix is the
generating function for this transformation.

Now that we have specified the Poisson brackets between the
parameters $(\mu_i,a_i)$, we can compute the Poisson
brackets between the matrix elements $V_{ij}$.
Remarkably, the latter can be written with the help
of a $d$-matrix depending on the dynamical variables.
\proclaim Proposition.
Denote by $E_{ij}=\ket{i}\bra{j}$
the canonical basis of the $N\times N$ matrices. Put
$V=\sum_{ij} V_{ij} E_{ij}$. Then:
\debut
\Bigl\{V_1\ \x\ V_2\Bigr\} = \BBL d_{12}, V_1\BBR - \BBL d_{21},V_2\BBR
\label{EAf}\fin
where $d_{12}= \sum_{ij;kl} d_{ij;kl}E_{ij}\otimes E_{kl}$ with,
\debut
d_{ij;kl} = -\inv{8}\({\frac{\mu_i +\mu_j}{\mu_i-\mu_j} }\)
\({ V_{jk}\de_{il}+V_{jl}\de_{ik}+V_{ik}\de_{jl}+V_{il}\de_{jk}}\)
\label{EAg}\fin
\par \noindent
In Eq.(\ref{EAf}), we used the standard notation in which
the lower indices refer to the space on which the matrices
are acting and $d_{21}= P d_{12}P$ with $P$ the flip operator.
Eq.(\ref{EAf}) can be proved by computing
both sides and comparing them.

Alternatively, eq.(\ref{EAf}) can be written in a quadratic form.
Let $M_{12}= \sum_{ij;kl} M_{ij;kl}E_{ij}\otimes E_{kl}$ with,
$M_{ij;kl}=-\inv{8}\({\frac{\mu_i +\mu_j}{\mu_i-\mu_j}}\)
\de_{jk}\de_{il}$, and define
\debut
r_{12}&=& M_{12}-M_{21} - (M_{12}^{t_1}-M_{21}^{t_2}) \non \\
s_{12}&=& M_{12}+M_{21} - (M_{12}^{t_1}+M_{21}^{t_2}) \nonumber
\fin
Then, we have:
\debut
\Bigl\{V_1\ \x\ V_2\Bigr\} = r_{12} V_1V_2 + V_2 s_{12} V_1 - V_1 s_{12} V_2
 - V_1 V_2 r_{12} \label{EAh}
\fin
Examples of this type of quadratic brackets first appeared in
\cite{Maillet1985}. Finally, we recall that the existence of a
$d$-matrix
is equivalent to the property that the eigenvalues of the matrix
$V$ are Poisson commuting \cite{BaVia90}.

\section{The N-soliton solution as a N-body problem.}

 From the definition of the sine-Gordon field in terms of the
$\tau$-functions, we see that only the eigenvalues $Q_i$ of the matrix $V$
are important. We have
\begin{eqnarray}
e^{-i\phi}=\prod_{i=1}^N \({ {1+Q_i\over 1-Q_i} }\)
\label{EphQ}
\end{eqnarray}
In this section, we rewrite the soliton dynamics in terms of the variables
$Q_i$. This leads to a relativistic generalization of the Calogero model,
first introduced by Ruijsenaars. We consider for definitness the evolution with
respect to the light cone variable $z_+=x+t$. Let $U$ be the matrix which
diagonalizes $V$:
\begin{eqnarray}
V=U^{-1}Q U
\label{matriceQ}
\end{eqnarray}
where $Q$ denotes the diagonal matrix ${\rm diag}(\,Q_i)$. We
define
\begin{eqnarray}
L= U \mu U^{-1}
\label{matriceL}
\end{eqnarray}
where $\mu$ denotes the diagonal matrix ${\rm diag}(\,\mu_i)$. As above,
$\mu_i$ are the
rapidities of the solitons. The matrix $L$ plays the role of a Lax
operator. Obviously, the quantities $tr(L^n)=\sum_{i=1}^N \mu_i^n $ are
conserved during the evolution of the solitons. Moreover, they are in
involution under the Poisson bracket (\ref{EAe}). Finally, we
have
\proclaim Proposition.
The ``time'' evolution of $L$ is given by a Lax equation
\begin{eqnarray}
\dot{L}=[ M,L],~~~M=\dot{U}U^{-1}
\nonumber
\end{eqnarray}
Here the dot means $\partial \over \partial z_+$.
Remarkably, $L$ and $M$ can be expressed in terms of the quantities
$Q_i$ and $\dot{Q}_i$:
\begin{eqnarray}
L_{ij}=2 {\sqrt{\dot{Q}_i \dot{Q}_j} \over Q_i +Q_j},
\quad {\rm and} \quad
M_{ij}&=&{\sqrt{\dot{Q}_i \dot{Q}_j} \over Q_i -Q_j}
(1-\delta_{ij})
\label{LetM}
\end{eqnarray}
\par \noindent
{\it Proof.} We start from the relation
\begin{eqnarray}
\mu V + V \mu = 2 |e>< e|
\nonumber
\end{eqnarray}
where $|e>$ is the column vector with components $e_i =\sqrt{X_i \mu_i}$. This
relation remains true when we go to the basis where $V$ is diagonal. Using
the definitions eqs.(\ref{matriceQ},\ref{matriceL}), we get
\begin{eqnarray}
Q L + L Q = 2 |\tilde{e}> < \tilde{e}|,~~~{\rm with}~~~|\tilde{e}> = U\,|e>
\nonumber
\end{eqnarray}
Since $Q$ is diagonal, we immediatly obtain from this relation
\begin{eqnarray}
L_{ij}= 2 {\tilde{e}_i \tilde{e}_j \over Q_i +Q_j}
\nonumber
\end{eqnarray}
Next, we remark that ($\dot{\sqrt{X_i}}= \mu_i \sqrt{X}_i$)
\begin{eqnarray}
\dot{V} = \mu V + V \mu = |e> < e|= U^{-1}\left( \dot{Q} + [Q, M]
\right) U
\nonumber
\end{eqnarray}
Multiplying this equation on the right by $U^{-1}$ and by $U$
on the left, and using the definition of $\ket{\tilde e}$, we get
$|\tilde{e}> < \tilde{e}| = \dot{Q} + [Q, M]$.
In components, this reads
\begin{eqnarray}
\tilde{e}_i \tilde{e}_j = \dot{Q}_i \delta_{ij} +(Q_i -Q_j) M_{ij}
\nonumber
\end{eqnarray}
If $i=j$, we find $\tilde{e}_i = \sqrt{\dot{Q}_i}$, and if $i\neq j$, we
find the value of $M_{ij}$ in terms of $Q_i,\;\dot{Q}_i$. This
completely determines the matrix $M$ since we can always ensure
that $M$ is antisymmetric by a suitable choice of basis.
This ends the proof.

\section{A canonical transformation.}

We now come to the description of the Poisson structure in terms of the
variables $Q_i,\, \dot{Q}_i$. Instead of the variables $\dot{Q}_i$, let us
introduce the variables $\rho_i = \dot{Q}_i/Q_i$. The Lax matrix becomes
\begin{eqnarray}
L_{ij}=2{\sqrt{Q_i Q_j}\over Q_i +Q_j} \sqrt{\rho_i \rho_j}
\nonumber
\end{eqnarray}
Comparing with eq.(\ref{EAc}), we see that $L$ has exactly the same form as
$V$ with the change of variables $(\mu_i,a_i) \to ( Q_i, \rho_i)$. This
symmetry extends at the level of the symplectic structure. We
have
\proclaim Proposition.
The transformation $ (\mu_i, a_i) \to (Q_i, \rho_i)$
is a symplectic transformation.
\par
\noindent {\it Proof.} We want to show that
\begin{eqnarray}
\omega &=&\sum_{i=1}^N \frac{\de a_i}{a_i}\wedge \frac{\de \mu_i}{\mu_i}
+ \sum_{i<j} \({\frac{4\mu_i\mu_j}{\mu_i^2-\mu_j^2} }\)
\frac{\de \mu_i}{\mu_i}\wedge \frac{\de \mu_j}{\mu_j}
\label{ECa} \\
&=& \sum_{i=1}^N \frac{\de \rho_i}{\rho_i}\wedge \frac{\de Q_i}{Q_i}
+ \sum_{i<j} \({\frac{4Q_iQ_j}{Q_i^2-Q_j^2} }\)
\frac{\de Q_i}{Q_i}\wedge \frac{\de Q_j}{Q_j}
\nonumber
\end{eqnarray}
Let us take as independent variables the pairs $(Q_i,\mu_i)$,
and expand both lines of eq.(\ref{ECa}) in these variables.
The equality of the two-forms then requires that:
\debut
\La_{ij}&=& \left( {\mu_j \over a_i} {\partial a_i \over \partial \mu_j}
-{\mu_i \over a_j} {\partial a_j \over \partial \mu_i} +{4\mu_i \mu_j \over
\mu_i^2 -\mu_j^2}\right) = 0 \label{ECb}\\
\hat \La_{ij}&=& \left( {Q_j \over \rho_i} {\partial \rho_i \over \partial Q_j}
-{Q_i \over \rho_j} {\partial \rho_j \over \partial Q_i} +{4Q_i Q_j \over
Q_i^2 -Q_j^2}\right) = 0 \nonumber\\
& &\frac{Q_j}{a_i}\frac{\d a_i}{\d Q_j} =
\frac{\mu_i}{\rho_j}\frac{\d \rho_j}{\d \mu_i} \nonumber
\fin
Let us first prove the third identity. By differentiating at
$\mu_i$ fixed
the matrix $V$ written in either forms eq.(\ref{EAc}) or
eq.(\ref{matriceQ}), we obtain:
\debut
d Q+\BBL Q, d U\, U^{-1}\BBR = U\, d V\, U^{-1}
=\half \({ (U\, a^{-1}da\, U^{-1})\, Q + Q\, (U\, a^{-1}da\, U^{-1})}\)
\nonumber
\fin
with $a$ the diagonal matrix ${\rm diag}(a_i)$. Using the fact that
both $Q$ and $a$ are diagonal matrices, we derive that:
\debut
\frac{dQ_i}{Q_i} = \sum_{j=1}^N U_{ij} \frac{da_j}{a_j} U^{-1}_{ji}
\nonumber
\fin
or equivalently,
\debut
\sum_{j=1}^N U_{ij} U^{-1}_{ji}
\({ \frac{Q_k}{a_j} \frac{\d a_j}{\d Q_k} }\) = \de_{ik}
\nonumber
\fin
Introducing the matrices ${\cal M}_{ij}=U_{ij}U_{ji}^{-1}$ and
${\cal A}_{jk}={Q_k\over a_j} {\partial a_j \over \partial Q_k}$, we can
rewrite this equation in matrix form:
\begin{eqnarray}
{\cal M A}= {\rm Id}
\label{ECk}
\end{eqnarray}

Similarly, using the expression of $L$ in terms of $\rho_i$ and $Q_j$,
we find, by differentiating the relation $L= U \mu U^{-1}$
at $Q_j$ fixed, that:
\begin{eqnarray}
{\cal R M}={\rm Id}
\label{ECl}
\end{eqnarray}
where ${\cal R}_{kj}= {\mu_k \over \rho_j} {\partial \rho_j \over
\partial
\mu_k }$.  Comparing eqs. (\ref{ECk}) anf (\ref{ECl}) proves the third
relation in (\ref{ECb}) since the inverse of ${\cal M}$ is
unique.

Let us now prove that $\La_{ij}=0$.
Expanding the first line of eq.(\ref{ECa}) in the variables
$(\mu_i,Q_i)$ we have
\debut
\omega=\sum_{ij} {\cal A}_{ji}
 {\delta Q_i \over Q_i} \wedge {\delta \mu_j \over \mu_j}
+ \sum_{i<j} \La_{ij} {\delta \mu_i \over \mu_i} \wedge {\delta \mu_j \over
\mu_j}
\nonumber
\fin
The Poisson brackets $\{Q_i,Q_j\}$ can be computed by inverting
the symplectic form. We find
\begin{eqnarray}
\{Q_i,Q_j\}= \left( {\cal A}^{-1} \Lambda \;^t{\cal
A}^{-1}\right)_{ij}
\nonumber
\end{eqnarray}
We know from the existence of the $d$-matrix (\ref{EAf}), that
these Poisson brackets vanish. Therefore ${\Lambda}_{ij} =0$.
The relations $\Lambda_{ij}=0$ are functional relations
between the variables $\mu_i, a_i$ at $Q_i$ fixed. Since the variables
$(\mu_i, a_i)$ and $(Q_i, \rho_i)$ play a completely symmetric
role, the conditions $\hat \La_{ij}=0$ are also satisfied.
This ends the proof that the transformation $(\mu_i,a_i)
\to (Q_i,\rho_i)$ is symplectic.

Explicitly, the Poisson brackets between $(Q_i,\rho_i)$ are:
\debut
\bigl\{Q_i\ ,\ Q_j \bigr\} &=& 0 \non\\
\bigl\{\rho_i\ ,\ Q_j \bigr\} &=& Q_j \rho_i \de_{ij}\non\\
\bigl\{\rho_i\ ,\ \rho_j\bigr\} &=&
- \({ \frac{4Q_iQ_j}{Q_i^2-Q_j^2}}\) \rho_i \rho_j \nonumber
\fin
It will be convenient to introduce a new set of
variables $p_i$, which are conjugated to the
coordinates $Q_i$.  We define them  by:
\debut
\rho_i =\exp(p_i) \prod_{k \not= i}\({ \frac{Q_k+Q_i}{Q_k-Q_i}}\)
\label{ECz}
\fin
The Poisson brackets are then canonical:
\debut
\bigl\{Q_i\ ,\ Q_j \bigr\} &=&
\bigl\{p_i\ ,\ p_j \bigr\} = 0 \non\\
\bigl\{p_i\ ,\ Q_j \bigr\} &=& Q_j \de_{ij}
\nonumber
\fin

\section{Hamiltonians and equations of motion.}

The result of the previous section has an immediate consequence.
The Poisson brackets of the Lax matrix $L$, given in eq.(\ref{LetM}),
are quadratic with a dynamical $R$-matrix:
\debut
\bigl\{L_1\ \x\ L_2\bigr\} = R_{12}L_1L_2 + L_2 S_{12} L_1
 - L_1 S_{12} L_2 - L_1 L_2 R_{12}
\label{Ecata}
\fin
where $R_{12}$ and $S_{12}$ are defined as $r_{12}$ and
$s_{12}$ in section 2, but with $\mu_i$ changed into $Q_i$.
This follows directly from the fact that
the transformation $(\mu_i,a_i)$ to $(Q_i,\rho_i)$ is
symplectic, and the fact that the Lax matrix $L$
is identical to the matrix $V$ but with $(\mu_i,a_i)$
changed into $(Q_i,\rho_i)$

This implies that the function $\CT(x)={\rm det}(1+xL)$
is a generating function of commuting Hamiltonians:
\debut
\bigl\{ \CT(x)\ ,\ \CT(y) \bigr\} = 0
\nonumber
\fin
Expanding it in power of $x$, we find:
\debut
\CT(x)&=& 1 + \sum_{p=1}^N x^p H_p\label{ECw}\\
&=&  1 + \sum_{p=1}^N x^p\ \sum_{k_1<\cdots<k_p}
\rho_{\rho_{k_1}}\cdots\rho_{k_p}
\prod_{k_i<k_j}\({\frac{Q_{k_i}-Q_{k_j}}{Q_{k_i}+Q_{k_j}}}\)^2
\nonumber
\fin
As is well known, the existence of a $R$-matrix ensures
that all the flow generated by these  Hamiltonians
admit a Lax pair formulation.
In particular, the Hamiltonian generating the evolution
in the light cone coordinate $z_+$ is $H_+={\rm tr}(L)$:
\debut
H_+= \sum_{j=1}^N \rho_j
 = \sum_{j=1}^N \exp(p_j)\ \prod_{k \not= j}\({ \frac{Q_j+Q_k}{Q_j-Q_k}}\)
\label{EDa}
\fin
The evolution in the other light cone coordinate $z_-$
is generated by the inverse of the Lax matrix. Its
Hamiltonian is $H_-={\rm tr}(L^{-1})$. Using
${\rm tr}(L^{-1})=H_{N-1}/{\rm det}(L)$, we find:
\debut
H_-= \sum_{j=1}^N \rho_j^{-1} \prod_{k\not= j}\({ \frac{Q_j+Q_k}{Q_j-Q_k}}\)^2
= \sum_{j=1}^N \exp(-p_j)\ \prod_{k \not= j}\({ \frac{Q_j+Q_k}{Q_j-Q_k}}\)
\label{EDb}
\fin
Notice that one goes from $H_+$ to $H_-$ by changing the sign
of $p_j$.

The equations of motion follow from these Hamiltonians.
For the light cone coordinate $z_+$, we have
\debut
\dot{Q_i} &=& \rho_i Q_i \non\\
\dot{\rho_i} &=& \sum_{k\not= i}
\({\frac{4Q_iQ_k}{Q_i^2-Q_k^2} }\) \rho_i\rho_k \non
\fin
where the dot still denotes $\frac{\d}{\d z_+}$.
The flow is the $z_-$ direction is very similar:
one just formally changes the dot by the derivative
$\frac{\d}{\d z_-}$. These are the equations of motion
written by Ruijsenaars \cite{Ruis}.

To write explicitly the equations of motion for the
(anti-) solitons and the breathers, we have
to disentangle the reality conditions on $(Q_i,p_i)$.
For the sine-Gordon field (\ref{EphQ}) and the Hamiltonians
(\ref{EDa},\ref{EDb}) to be
real, the coordinates $(Q_i,p_i)$ have to come by
pairs $(j,\bar j)$, with $Q_{\bar j}=-\bar Q_j$ and $p_{\bar
j}=\bar p_j$.
The case $j=\bar j$ corresponds to a soliton or an
antisoliton, and the case $j\not= \bar j$ to a breather.
Therefore, we introduce new coordinates $q_j$ by $Q_j=ie^{q_j}$
such that:
\begin{eqnarray}
{\rm Im}~q_s=0  && {\rm for}~ s~ {\rm a~ soliton} \nonumber \\
{\rm Im}~q_{\bar s}=\pi  && {\rm for}~ \bar s~ {\rm an~ antisoliton} \nonumber
\\
q_{\bar b}= {\bar q_{b}}  && {\rm for}~ b~
{\rm a~ breather} \nonumber
\end{eqnarray}
Similarly the momenta $p_s$ and $p_{\bar s}$ are real and
$p_b$ is complex with $p_{\bar b}={\bar p_b}$.
In these coordinates, the Hamiltonians $H_\pm$ become:
\begin{eqnarray}
H_\pm = \sum_j e^{\pm p_j}\ \prod_{k\not= j}
\coth\({\frac{q_j-q_k}{2} }\) \nonumber
\end{eqnarray}
The equations of motion for the flow generated by $H_+$ read:
\begin{eqnarray}
\rho_i &=& \dot{q_i} = e^{p_i} \prod_{k\not= i}
\coth\({\frac{q_i-q_k}{2} }\) \nonumber\\
\ddot{q_i} &=& \sum_{k\not= i}
\frac{2\dot{q_i}\dot{q_k}}{\sinh(q_i-q_k)}
\nonumber
\end{eqnarray}
By construction, these evolution equations are linearized
in the variables $(\mu_i,a_i)$.

\end{document}